\documentclass[aps,prb,showpacs,floatfix,reprint]{revtex4-1}
\usepackage[latin1]{inputenc}
\usepackage{amssymb}
\usepackage{verbatim}
\usepackage[usenames]{color}
\usepackage{graphicx}
\usepackage{epstopdf}

\usepackage{graphicx}

\begin{document}
\title{The effect of the buffer layer coupling on the lattice parameter of epitaxial graphene on SiC(0001)}
\author{T. Schumann}
\author{M. Dubslaff}
\author{M. H. Oliveira Jr.}\altaffiliation{Current Address: Departamento de Física, ICEx, Universidade Federal de Minas Gerais-UFMG, C.P. 702, 31270-901, Belo Horizonte, MG, Brazil}
\author{M. Hanke}
\author{J. M. J. Lopes}
\email{Corresponding author: lopes@pdi-berlin.de}
\author{H. Riechert}
\affiliation{Paul-Drude-Institut für Festk\"orperelektronik, Hausvogteiplatz 5--7, 10117 Berlin, Germany}

\begin{abstract}
Grazing incidence X-ray diffraction (GID) was employed to probe the structure of atomically thin carbon layers on SiC(0001): a so-called buffer layer (BL) with a $6(\sqrt{3}\times\sqrt{3})$R30$^\circ$ periodicity, a monolayer graphene (MLG) on top of the BL, and a bilayer graphene (BLG). The GID analysis was complemented by Raman spectroscopy. The lattice parameter of each layer was measured with high precision by GID. The BL possesses a different lattice parameter and corrugation when it is uncovered or beneath MLG. Our results demonstrate that the interfacial BL is the main responsible for the strain in MLG. By promoting its decoupling from the substrate via intercalation, it turns into graphene, leading to a simultaneous relaxation of the MLG and formation of a quasi-free-standing BLG. 
\end{abstract}

\date{\today}



\maketitle

Graphene with its broad range of superlative properties is promising for several applications, and thus anticipated to play a major role in future technologies.\cite{NovoselovNature2012} The practical utilization of this material will require the development of scalable processes aiming at its precise synthesis.\cite{NovoselovNature2012,BonaccorsoMT2012} As an example, for the production of graphene-based electronic devices, large-area growth of layers offering high structural and electronic quality directly on (semi-)insulating substrates is of great advantage as it will avoid transfer processes that may otherwise degrade their properties. This has currently been pursued by different groups which employed synthesis methods such as chemical vapor deposition,\cite{StrupinskiNL2011,HwangACS2013} molecular beam epitaxy,\cite{OliveiraCarbon2013b,MoreauAPL2010} as well as graphitization of SiC surfaces.\cite{DeHeerPNAS2011,EmtsevNMat2009,VirojanadaraPRB2008,OliveiraAPL2011,OstlerPRB2013,ColettiAPL2011} The latter technique makes use of high temperature annealing (usually above 1400\,$^\circ$C) to sublimate Si atoms and create a C-rich SiC surface where graphene is formed. 

Epitaxial graphene can be prepared on polar [(0001) and (000$\overline{1}$)] \cite{DeHeerPNAS2011,EmtsevNMat2009,VirojanadaraPRB2008,OliveiraAPL2011} and non-polar [(11$\overline{2}$0) and (1$\overline{1}$00)] \cite{OstlerPRB2013} faces of hexagonal SiC, as well as on the (111) surface of cubic SiC. \cite{ColettiAPL2011} Growth on the Si-terminated (0001) face is certainly the most investigated case as it has an important aspect: it allows for the preparation of high-quality monolayer graphene (MLG) which continuously covers surface terraces and steps in a carpet-like manner, with only small fractions of bi- or few-layer graphene existing close to step edges.\cite{EmtsevNMat2009,OliveiraAPL2011,FirstMRS2010} The interfacial layer between graphene and SiC is also one-atom thick and is often referred to as buffer layer (BL). Although there has been an intense debate about its atomic structure, \cite{ChenSS2005,WengAPL2010,QiPRL2010,GolerC2013,EmtsevPRB2008} it is currently almost a consensus that the BL is a $6(\sqrt{3}\times \sqrt{3})$R30$^\circ$ surface reconstruction of SiC exhibiting a graphene-like honeycomb lattice with partial sp$^3$-hybridization. \cite{GolerC2013,EmtsevPRB2008} It is also well accepted that the structural (and electronic) properties of the MLG are influenced by the existence of the BL. \cite{SchmidtPRB2011,RisteinPRL2012} Nevertheless, it is not completely understood how the structure and morphology of the BL changes during the growth of epitaxial graphene and/or due to post-growth processes (e.g. decoupling from the substrate by intercalation) \cite{RiedlPRL2009,OliveiraCarbon2013a} and, most importantly, how such modifications will finally affect the uppermost MLG. In order to shine further light on this issue, we have studied different types of carbon coverages on SiC(0001) samples using Raman scattering spectroscopy and grazing-incidence X-ray diffraction (GID). The use of the latter for the characterization of multilayer graphene films on SiC has been demonstrated. \cite{GoncalvesNT2012} However, no emphasis was placed on the BL, or on high-precision measurements of a single graphene layer. GID allowed us to measure the in-plane lattice parameters of graphene (mono- and bilayer) and BL on SiC(0001) with very high precision. Based on this, information about the average strain level in each atomic layer could be gained, which agrees with Raman results. It is observed that the BL possesses different lattice parameter and corrugation for the cases when it is uncovered or covered by a MLG. It is also revealed that the interfacial BL is indeed the main agent responsible for the strain normally measured in MLG on SiC(0001). The present results are of general relevance as they show that GID is a powerful tool for precise structural studies of purely 2D atomic crystals. 
	
\begin{figure}[tb]
\includegraphics[width=\linewidth]{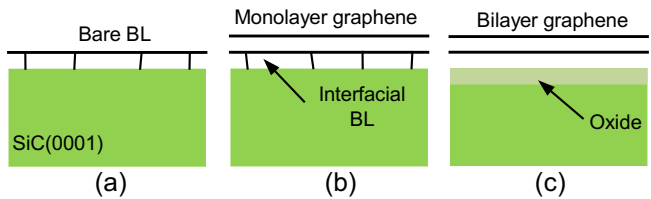}
\caption{Schematic structures (side view) of the samples investigated in the present study: (a) bare BL; (b) monolayer graphene on top of the BL; (c) bilayer graphene on top of an oxidized SiC surface (obtained by oxygen intercalation).} \label{Fig1}
\end{figure}

Three types of samples were investigated in this work, as illustrated in Figure 1. A bare BL (Fig.~\ref{Fig1}a), MLG (on top of the BL - Fig.~\ref{Fig1}b), as well as bilayer graphene (BLG) (Fig.~\ref{Fig1}c), which were all prepared on n-type 6H-SiC(0001). The substrates were chemically cleaned and hydrogen-etched using a standard procedure. \cite{OliveiraAPL2011} The bare BL sample (Fig.~\ref{Fig1}a) was grown at a temperature of 1400\,$^\circ$C for 15\,min in an Ar atmosphere of 900\,mbar and a flow rate of 100\,sccm. The MLG sample (Fig.~\ref{Fig1}b) was prepared at a temperature of 1600\,$^\circ$C for 15\,min in an Ar atmosphere of 900\,mbar and a flow rate of 500\,sccm. Note that due to the layer-by-layer growth (from below) of graphene on SiC(0001), \cite{TanakaPRB2010} the first carbon layer formed during the graphitization process is in fact a bare BL. It will turn into a purely sp$^2$-hybridized layer (i.e. graphene) only when a second BL is formed underneath it. This is because such a process eliminates the sp$^3$-bonds and thus promotes its detachment from the SiC. Hence, the underlying layer becomes the new BL, while the former (bare BL) converts into a graphene monolayer. The BLG sample (Fig.~\ref{Fig1}c) was synthesized in two steps: i) MLG was prepared as for the sample illustrated in Fig.~\ref{Fig1}b; ii) bilayer formation was achieved by oxygen intercalation upon annealing in air for 40 min at 600\,$^\circ$C. During this process, oxygen-containing species intercalate underneath the MLG/BL and oxidize the SiC surface. The BLG is created as this process decouples the BL from the substrate, and turns into a graphene layer. More details about the O$_2$ intercalation process upon air annealing are reported elsewhere.\cite{OliveiraCarbon2013a}

\begin{figure}[tb]
\includegraphics[width=\linewidth]{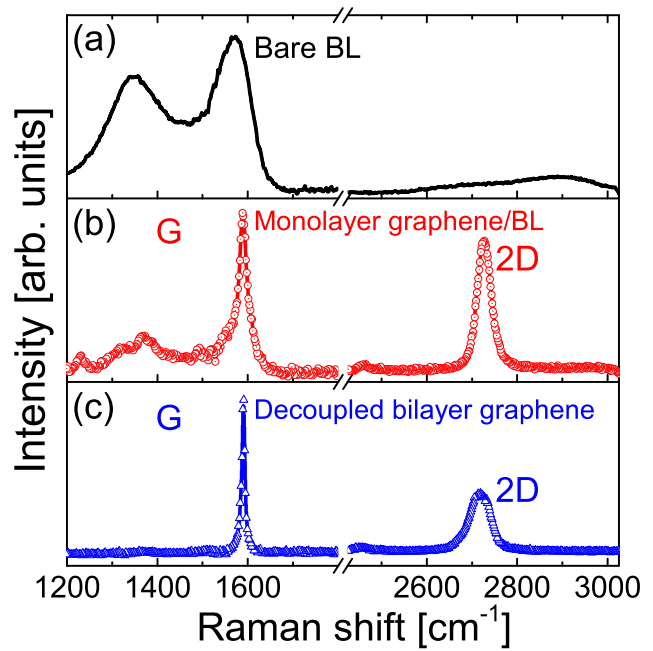}
\caption{Raman spectra of the three samples investigated in the present study: (a) bare BL; (b) monolayer graphene on top of the BL; (c) bilayer graphene on top of an oxidized SiC surface (obtained by oxygen intercalation upon air annealing).} \label{Fig2}
\end{figure}

Raman spectra of the three different samples are shown in Fig.~\ref{Fig2}. They were recorded using an excitation wavelength of 482.5\,nm with a spatial resolution of 1\,$\mu$m. The measurements were performed exclusively on surface terraces to avoid contributions from fewlayer graphene at the step edge regions. \cite{EmtsevNMat2009,OliveiraAPL2011} The spectrum recorded for the bare BL (see Fig.~\ref{Fig2}a) exhibits two intense and broad bands in the spectral region of 1200\,--\,1660\,cm$^{-1}$ and a low-intensity modulated bump between 2540 and 3000\,cm$^{-1}$. Well-defined G and 2D peaks, which are usually measured for graphene, \cite{FerrariPRL2006} are not seen in the spectrum. This is because the BL possesses a phonon dispersion which is substantially different from the one of graphene. \cite{FrommNJP2013} For the MLG/BL system, a Raman spectrum showing intense G and 2D peaks is measured (see Fig.~\ref{Fig2}b). The 2D peak can be well fitted by a single Lorentzian, as expected for a single layer. The G and 2D peaks are positioned at 1581($\pm$5)\,cm$^{-1}$ and 2724($\pm$10)\,cm$^{-1}$, respectively. Based on the position of the Raman peaks, \cite{ZabelNL2012} an average (compressive) strain of $\sim$0.2\,\%\ was estimated for the MLG. The position of 2D peak was utilized for this purpose, since it is only marginally affected by the doping, present in the investigated samples.\cite{DasNT2008,DasPRB2009} The broad spectral features existing from 1200\,cm$^{-1}$ close to the G peak's left shoulder originate from the underlying BL. \cite{FrommNJP2013} The Raman spectrum collected after air annealing (see Fig.~\ref{Fig2}c) shows features of quasi-free-standing BLG. The 2D peak can be fitted by four Lorentzians positioned at 2673\,cm$^{-1}$, 2688\,cm$^{-1}$, 2706\,cm$^{-1}$, and 2738\,cm$^{-1}$ (fittings not shown). This coincides well with the values obtained by Malard \textit{et al.} \cite{MalardPRB2007} for exfoliated BLG on SiO$_2$ (taken the employed excitation energy into account), which shows that the original compressive strain present in the MLG is released after BL decoupling. This in turn suggests that the existence of the BL at the interface is certainly one of the main factors contributing to the compressive strain observed in the graphene layer. \cite{OliveiraAPL2011,SchmidtPRB2011} Another evidence for the formation of BLG is that the BL-related Raman features are absent in the spectrum. Note that, as previously reported,\cite{OliveiraCarbon2013a} the absence of a D peak proves that air annealing process does not lead to defect formation in the graphene structure.

The structure of the samples was further investigated by GID. The measurements were performed at the beamline ID10 of the European Synchrotron Radiation Facility (ESRF) in Grenoble. The primary beam energy was 10\,keV with an intensity of 10$^{14}$ counts per second (cps) and a spot size on the sample of 100\,$\mu$m$\times$1\,mm. The angle of incidence was set to 0.15\,$^\circ$, which is below the angle of total external reflectance (0.21\,$^\circ$ for SiC and 10\,keV), to minimize the intensity obtained from the substrate. Using this technique, the lattices planes orthogonal to the sample surface normal are analyzed by diffraction, and information about in-plane lattice parameter and orientation can be acquired. Figure 3 illustrates a reciprocal space map (RSM) obtained by combining angular and radial scans of GID measurements performed for the MLG/BL on SiC(0001). Two SiC-related reflections are present, the SiC(2$\overline{11}$0) and the (quasi-forbidden) SiC(2$\overline{2}$00). Two graphene-related reflections are also observed, assigned as G(10$\overline{1}$0) and G(11$\overline{2}$0). The appearance of these isolated reflections reveals that the layer possesses a single orientation with respect to the substrate. The graphene lattice is rotated by 30\,$^\circ$ with respect to the SiC, since the equivalent SiC (2$\overline{11}$0) and graphene (11$\overline{2}$0) reflections are rotated by this angle relative to each other. Very similar maps (not shown) were obtained for the other two samples investigated here. 

In order to measure possible modifications in the lattice parameters of the three samples, line scans over the G(11$\overline{2}$0) reflection were performed with higher resolution along the radial direction q$_{\text{r}}$, as shown in Fig.~\ref{Fig4}. The central position of the peaks was determined by fitting Gaussians to the curves. The measurements reveal clear differences between the samples. The bare BL shows a lattice parameter of a\,=\,2.467\,\AA. The G(11$\overline{2}$0) reflection of the MLG/BL sample shows splitting, and can be well fitted with two Gaussian peaks, centered at 2.456\,\AA\ and 2.463\,\AA. The reflection obtained from the BLG consists of a single Gaussian, resulting in a lattice parameter of a\,=\,2.460\,\AA. Note that the upper limit for error is estimated to be $\sim$0.001\,\AA, based on the fitting error, the energy resolution of the primary beam, the accuracy of the motors which move the sample and the detector, and the alignment of the substrate-related peaks to literature values. Table \ref{tab:Table1} summarizes the lattice parameter measured for each sample.

\begin{table}[b]
	\centering
		\begin{tabular}{|c|c|}
			\hline
			\bf Sample& \bf a\,[\AA]\\
			\hline
			Bare BL & 2.467\\
			\hline
			Monolayer graphene/BL & 2.456/2.463\\
			\hline
			Decoupled bilayer graphene & 2.460\\
			\hline
		\end{tabular}
	\caption{Lattice parameter \textbf{a} obtained by GID for the samples investigated here.}
	\label{tab:Table1}
\end{table}

\begin{figure}[tb]
\includegraphics[width=\linewidth]{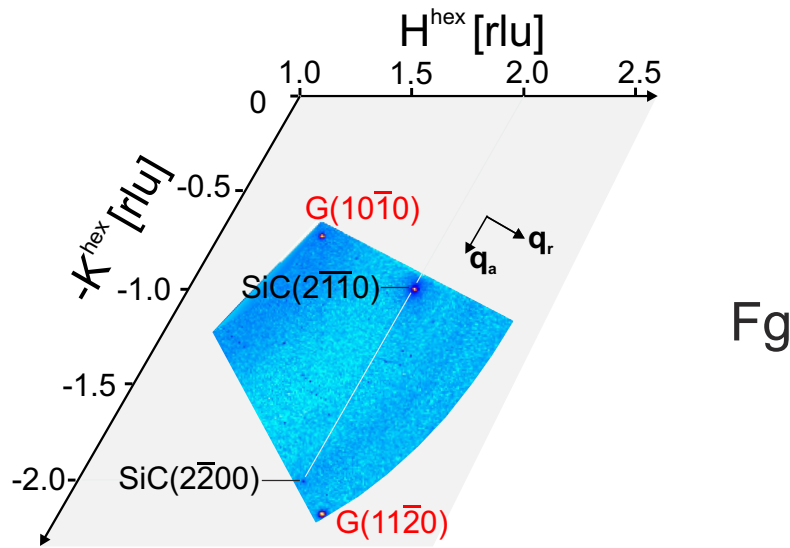}
\caption{Reciprocal space map of monolayer graphene/BL on SiC(0001). The axes are scaled with the reciprocal lattice units (rlu) of SiC, q$_{\text{a}}$ and q$_{\text{r}}$ mark the radial and angular directions in the RSM.} \label{Fig3}
\end{figure}

\begin{figure}[tb]
\includegraphics[width=\linewidth]{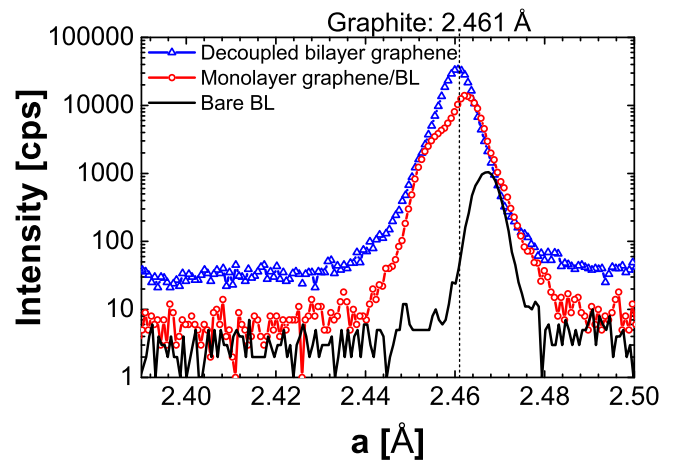}
\caption{Linescans through the G(11$\overline{2}$0) reflection along q$_{\text{r}}$, transformed to real-space, performed for the bare BL, monolayer graphene/BL, and bilayer graphene samples. The dotted line indicates the value for the lattice parameter of graphite. \cite{HattabNL2012} The x-axis is scaled to the lattice parameter of graphene.} \label{Fig4}
\end{figure}

The lattice parameter for the bare BL is $\sim$0.24\,\%\ larger in comparison to that of graphite (a\,=\,2.461\,\AA). \cite{HattabNL2012} The larger lattice parameter of the BL is a result of its strong bonding to the substrate due to the sp$^3$-hybridization of $\sim$1/3 of the C-atoms. \cite{EmtsevPRB2008} These sp$^3$-bonds will likely affect the inter-atomic distances in the BL lattice and consequently the average in-plane lattice parameter. In fact, it has been shown that the bond length between sp$^3$- and sp$^2$- hybridized carbon atoms are approximately 3\,\%\ longer than that between two sp$^2$-hybridized atoms. \cite{BrownTFS1959} Furthermore, the partial sp$^3$-bonding to the substrate leads to a ($6\times6$) long-range corrugation in the BL (as imaged by scanning tunneling microscopy - STM). \cite{GolerC2013} We could roughly estimate the out-of-plane height for this corrugation (see Supplemental Material).\cite{suppmat} Assuming that an ideally flat BL on SiC would adapt graphene's lattice parameter of 2.461\,\AA, an average corrugation angle of $\sim$4$^\circ$ to the surface is required to obtain the measured lattice parameter of the BL. Based on it and on STM literature data which shows that the buckled region is extended over the side of the ($6\times6$) cell, a corrugation of $\sim$0.52\,\AA\ was obtained. This value is similar to what was measured by STM. \cite{GolerC2013} It is important to mention that the enlarged lattice parameter measured by GID evidences that the uncovered BL possesses a graphene-like hexagonal arrangement. The existence of pentagons, heptagons, and even octagons (due to either inclusion of extra carbon atoms or formation of vacancies) is expected to lower the average bond length and thus lattice parameter. \cite{NemecPRL2013,SchumannNJP2013} 
	
The doublet shape of the G(11$\overline{2}$0) reflection obtained from the MLG/BL sample is likely related to the different bonding characteristics of the two layers. While the uppermost MLG is purely sp$^2$-hybridized, the interfacial BL is expected to have the same density of sp$^3$-bonds as for the case when it is uncovered. \cite{EmtsevPRB2008} Therefore, the fact that they exhibit different in-plane lattice parameters (2.456 and 2.463\,\AA\ for the MLG and the BL, respectively) is not totally unexpected. GID reveals other two interesting aspects. The first one is related to the MLG. Its lattice parameter is $\sim$0.22\,\%\ smaller in comparison to graphite. The magnitude and type of strain is in agreement with what was estimated by Raman spectroscopy ($\epsilon$\,=\,$\sim$0.2\,\%). It has been proposed that the compressive strain may arise upon sample cooling due to the different coefficients of thermal expansion for graphene and SiC. \cite{FerralisPRL2008} While graphene's coefficient is negative, the one for SiC is positive. \cite{YoonNL2011} Hence, for the contraction of the graphene lattice to be caused by this phenomenon, the MLG must be strongly pinned to the underlying BL/SiC substrate. The origin of the pinning remains unknown. In principle, one cannot exclude the possibility that covalent bonds are formed between the graphene and the BL, especially at grain boundaries. However, no experimental evidences supporting this hypothesis have been reported so far. Another tentative explanation has been given by Ferralis \textit{et al.} \cite{FerralisPRB2011} They suggest that the highly corrugated potential in the substrate surface (which is indeed expected as the BL exhibits a semiconducting nature) \cite{EmtsevPRB2008} will promote a lateral pinning of the MLG, which will hinder tangential displacements for strain relaxation.
	
The second interesting aspect relates to the interfacial BL, which has a larger lattice parameter than that of graphite, similar to what was measured for the bare BL. However, the increase is smaller than in that case and amounts to only $\sim$0.08\,\%. The reason for the difference in the lattice parameters of the interfacial and the bare BL is not known yet, although one can speculate that it is associated with the different growth environments faced by the two layers. The bare BL grows at a SiC-Ar interface, while the interfacial BL grows underneath the former one. \cite{TanakaPRB2010} Also the existence of a graphene layer on top during sample cooling might be the reason for the smaller lattice parameter of the interfacial BL, e.g. due to a decrease in its long-range corrugation. Lauffer \textit{et al.} \cite{LaufferPRB2008} have observed (by STM) that the surface roughness of MLG/BL is lower than of the bare BL. Such reduction might be associated (at least to a certain extent) to the smoothing of the interfacial BL. \cite{deLimaPRB2013} The corrugation that we obtain based on the GID results is $\sim$0.29\,\AA, thus smaller than the value found for the bare BL ($\sim$0.52\,\AA). Finally, the fact that the bare and interfacial BL exhibit similar lattice parameters corroborates the mostly common interpretation in terms of structure, i.e. that both of them possess a periodic hexagonal structure like graphene. Should the $6(\sqrt{3}\times \sqrt{3})$R30$^\circ$ graphene-like structure not persist at the interface (as proposed for instance in Ref. \onlinecite{WengAPL2010}), the splitting of the G(11$\overline{2}$0) reflection for the MLG/BL structure would certainly not be observed.

The linescan over the G(11$\overline{2}$0) reflection for quasi-free-standing BLG contains a single peak centered at 2.460\,\AA. The difference relative to the value for graphite is $\sim$0.04\,\%. This very small compressive strain might also be caused by a pinning to the underlying SiO$_2$ due to effects related to surface potential, \cite{FerralisPRB2011} as previously discussed. This result is in agreement with the Raman findings [note that in that case we considered, as a reference for strain-free material, Raman data obtained for BLG flakes on SiO$_2$ (Ref. \onlinecite{MalardPRB2007})]. The single lattice parameter measured for both layers shows that, as a product of the oxygen intercalation, the interfacial BL decouples and slightly contracts, while (and as a consequence of it) the uppermost graphene expands. This unequivocally demonstrates that the interfacial BL with its strong bonding to the SiC is indeed responsible for the compressive strain generally measured in MLG on SiC(0001).  

In summary, we have used GID to investigate the structure of three atomically thin carbon films (bare BL, MLG/BL, and decoupled BLG) on SiC(0001). The GID analysis were complemented by Raman measurements. The in-plane lattice parameter of each atomic layer could be determined with high precision using GID. This permitted to gain information about their strain level. The results reveal that the BL possesses a lattice parameter that is larger than that of graphite. Interestingly, this value slightly decreases when the BL is located beneath a MLG, likely due to a lowering of the out-of-plane corrugation. Furthermore, our findings corroborate the mostly accepted interpretation for the structure of the BL, i.e. that it exhibits a graphene-like hexagonal lattice with part of the C atoms connected to the SiC substrate through sp$^3$-bonds. It also shows that the BL (as a semiconducting interfacial layer) is directly responsible for the strained nature of MLG. If decoupled from the substrate by oxygen intercalation, the BL turns into an almost strain-free graphene layer, allowing the MLG on top to move laterally and relax, ultimately forming a quasi-freestanding BLG structure.  

\begin{acknowledgments}
The authors thank A. Pr\"o\ss dorf for the critical reading of the manuscript; F. Grosse for fruitful discussions; C. Herrmann, M. H\"oricke and H.-P. Sch\"onherr for technical support; the ESRF for providing beamtime during project SI-2449 and Roberto Nervo for his support during the experiment.
\end{acknowledgments}

\bibliographystyle{apsrev}

\end{document}


\title{Supplementary material}
\author{T. Schumann}
\author{M. Dubslaff}
\author{M. H. Oliveira Jr.}\altaffiliation{Current Address: Departamento de Física, ICEx, Universidade Federal de Minas Gerais-UFMG, C.P. 702, 30123-970, Belo Horizonte, MG, Brazil}
\author{M. Hanke}
\author{J. M. J. Lopes}
\email{Corresponding author: lopes@pdi-berlin.de}
\author{H. Riechert}
\affiliation{Paul-Drude-Institut für Festk\"orperelektronik, Hausvogteiplatz 5--7, 10117 Berlin, Germany}

\maketitle

\section{Atomic force microscopy}

\begin{figure}[htb]
\includegraphics[width=\linewidth]{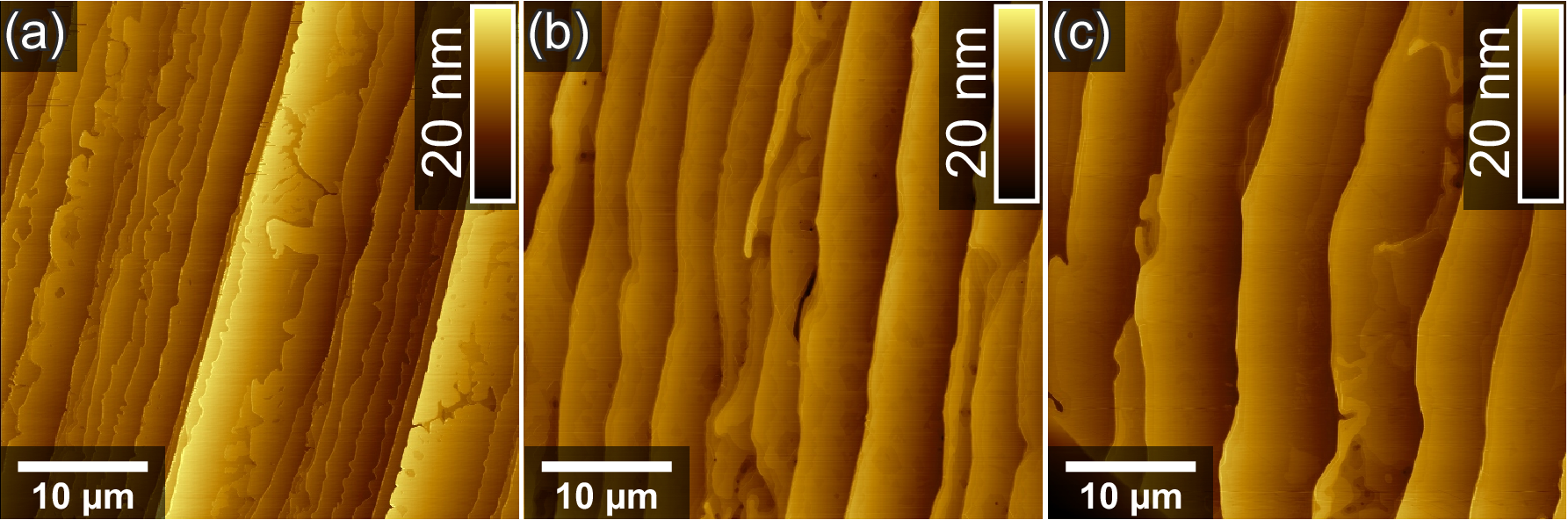}
\caption{Atomic force micrographs of (a) a buffer layer, (b) monolayer graphene, and (c) bilayer graphene.} \label{S1}
\end{figure}

AFM images from the surfaces of the buffer layer, the epitaxial monolayer graphene, and the quasi-freestanding bilayer graphene samples are depicted in Fig.~\ref{S1}. In all three samples, mostly atomically flat terraces with widths in the order of $\mu$m and step edges with heights of 5--10\,nm are present on the surface. Comparing Fig.~\ref{S1}~(b) and (c), the morphology does not change during the intercalation process. In particular no wrinkles appear on the surface, which could be related with a strain relieve process, as it is the case during growth of epitaxial graphene on C-face SiC \cite{SrivastavaJoPD2012}.

\section{Determination of corrugation via grazing incidence diffraction data}

\begin{figure}[htb]
\includegraphics[width=\linewidth]{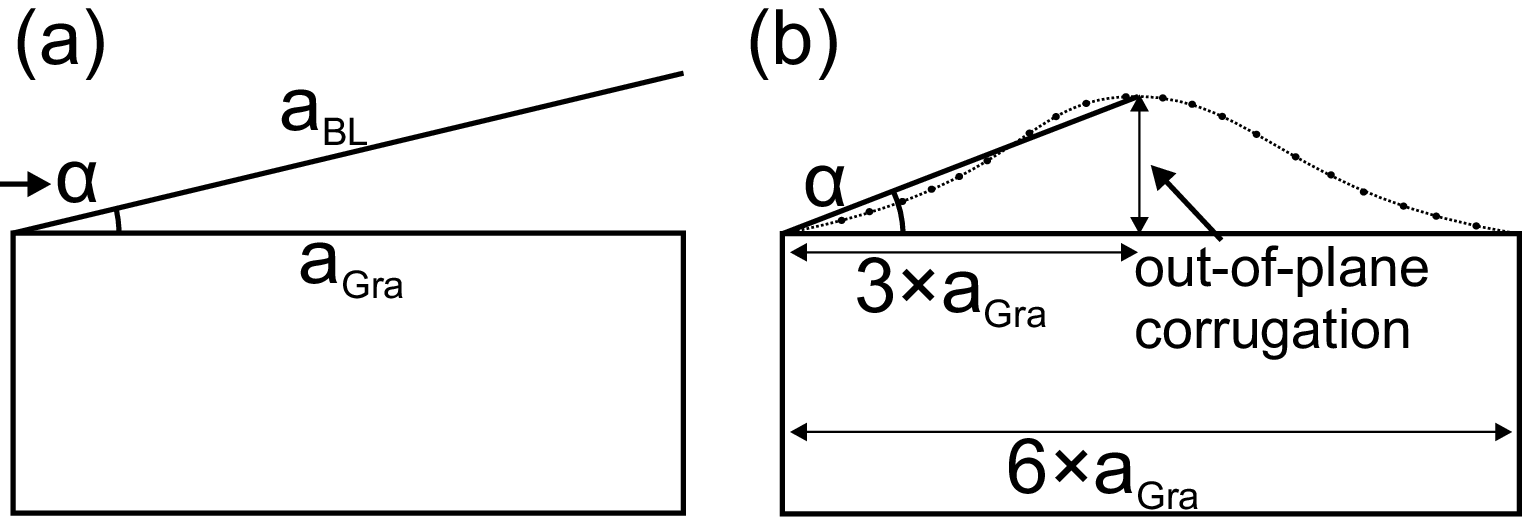}
\caption{Schematic model to determine the out-of-plane corrugation. See text for details}\label{S2}
\end{figure}

The (long-range) corrugation of the buffer layer was determined by applying the following model, as depicted in Fig.~\ref{S2}.

\begin{enumerate}
  \item We assume that a layer which is flat on the substrate surface would adapt the lattice constant of graphene, since the layers are commensurate. 
  \item The layer buckles up with such an angle $\alpha$, that the layers are in-plane commensurate (see Fig.\ref{S2}~(a)), hence  $\alpha = \sin{(\text{a}_{\text{Gra}}/\text{a}_{\text{BL}})}$.
  \item The periodicity of the long-range corrugation corresponds to the $(6\times 6)$ surface structure, as shown by scanning tunneling microscopy \cite{GolerC2013}. Therefore, the lateral  distance between \lq\lq valley\rq\rq\ and \lq\lq peak\rq\rq\ is $3\times \text{a}_{\text{Gra}}$ (see Fig.~\ref{S2}~(b)).
  \item Therefore, the out-of-plane corrugation of the BL is given by $3\times \text{a}_{\text{Gra}} \times \tan{\alpha}$.
\end{enumerate}

\bibliographystyle{apsrev}